\documentclass[12pt,thmsa]{article}
\usepackage{amsmath}
\usepackage{amssymb}
\usepackage{amsfonts}
\usepackage[T1]{fontenc}

\begin{document}

\title{The Geometry of Regular Shear-Free Null Geodesic Congruences, CR
functions and their Application to the Flat-Space Maxwell Equations}
\author{Carlos Kozameh$^{1}$ \and E.T. Newman$^{2}$ \and Gilberto
Silva-Ortigoza$^{3} $ \\
$^{1}$FaMaF, Univ. of Cordoba, \\
Cordoba, Argentina\\
$^{2}$Dept of Physics and Astronomy, \\
Univ. of Pittsburgh, \\
Pittsburgh, PA 15260, USA\\
$^{3}$Facultad de Ciencias F\'{\i}sico Matem\'{a}ticas \\
de la Universidad Aut\'{o}noma de Puebla, \\
Apartado Postal 1152, 72001,\\
Puebla, Pue., M\'{e}xico}
\date{6/15/07}
\maketitle

\begin{abstract}
We describe here what appears to be a new structure that is hidden in all
asymptotically vanishing $Maxwell$ fields possessing a non-vanishing total
charge. Though we are dealing with \textit{real} Maxwell fields on \textit{%
real} Minkowski space nevertheless, directly from the asymptotic field one
can extract a complex analytic world-line defined in complex Minkowski space
that gives a unified Lorentz invariant meaning to both the electric and
magnetic dipole moments. In some sense the world-line defines a `complex
center of charge' around which \textit{both} electric and magnetic dipole
moments vanish. \

The question of how and where does this complex world-line arise is one of
the two main subjects of this work. The other subject concerns what is known
in the mathematical literature as a CR structure. In GR, CR structures
naturally appear in the physical context of shear-free (or asymptotically
shear-free) null geodesic congruences in space-time. For us, the CR
structure is associated with the embedding of Penrose's \textit{real}
three-dimensional null infinity, $\mathfrak{I}^{+},$ as a surface in a two
complex dimensional space, $\mathrm{C}^{2}.$ It is this embedding, via a
complex function, (\textit{a CR function), }that\textit{\ }is\textit{\ }our
other area of interest. Specifically we are interested in the
`decomposition' of the CR function into its real and imaginary parts and the
physical information contained in this decomposition.
\end{abstract}

\section{Introduction}

The vacuum Maxwell equations have been with us for a long time and it would
appear unlikely that a potentially important or even interesting new
fundamental aspect of them would be found. Nevertheless that is our claim.

More specifically, we have found what appears to be a new structure hidden
in all asymptotically vanishing $Maxwell$ fields possessing a non-vanishing
total charge. Though we are dealing with \textit{real} Maxwell fields on
\textit{real} Minkowski space nevertheless, directly from the asymptotic
field one can extract a complex analytic world-line that is `living' in
complex Minkowski space that gives a unified Lorentz invariant meaning to
both the electric and magnetic dipole moments. In some sense the world-line
defines a complex center of charge around which \textit{both} electric and
magnetic dipole moments vanish. Normally both dipole moments are defined
either via charge and current integrals at any one time in a given Lorentz
frame $-$ or are extracted (also in a given Lorentz frame) from knowledge of
the asymptotic fields. Their values in another Lorentz frame are not
obtained from their previous values and thus they do not define Lorentzian
geometric quantities. We however will see that both dipole moments can be
found directly from the complex world-line in any Lorentz frame.

The question of how and where does this complex world-line arise is one of
the two main, but closely related, subjects of this work.

The other subject concerns a certain (pure) mathematical structure, known as
a CR structure, that has naturally appeared in the physical context of
shear-free (or asymptotically shear-free) null geodesic congruences in
space-time. For us, the CR structure is associated with the embedding of
Penrose's \textit{real} three-dimensional null infinity, $\mathfrak{I}^{+},$
as a surface in a two complex dimensional space, $\mathbf{C}^{2}.$ It is
this embedding, which involves the use of a complex function, \textit{a CR
function, }that\textit{\ }is\textit{\ }our other area of interest. More
specifically we are interested in the `decomposition' of the CR function
into its real and imaginary parts and the physical information contained in
this decomposition. We mention now and discuss later how every
asymptotically flat Maxwell field, with non-vanishing charge, induces a CR
structure on $\mathfrak{I}^{+}$.

It has been known for a long time in the relativity community that \textit{\
shear-free null geodesic congruences} play an important role in general
relativity\cite{RT,GS,RTM} and Maxwell theory\cite{IR,Sh.Free.Max}.
Penrose's twistor theory had its origins close to the study of such
congruences. Though it is difficult to argue from a priori knowledge that
this should be true, nevertheless from many examples and theoretical
discoveries, the importance does become easy to see. For example, many of
the most important exact solutions of the vacuum Einstein or
Einstein-Maxwell equations possess a degenerate principle null vector field
that is both geodesic and shear free, e.g., the Schwarzschild, the
Reissner-Nordstrom, the Kerr and Kerr-Newman metrics. In electrodynamics,
the Coulomb, the null\cite{IR} and the Lienard-Wiechert \{as well as the
complex Lienard-Wiechert\cite{Sh.Free.Max}\} Maxwell fields have a principle
null vector that is tangent to a shear free null geodesic congruence. From
these observations one was led to the more general issue of finding all
Einstein metrics that possess a principle null vector field that is
shear-free and geodesic. This in turn led to the discovery of the
algebraically special metrics and the beautiful Goldberg-Sachs theorem which
stated that the degenerate principal null vectors \textit{for Einstein
metrics} are always geodesic and shear free. In the study of such metrics
one of the very pretty mathematical discoveries was the automatic existence
of an associated three-dimensional CR structure\cite{LNT,LNT2,LN,Taf,Tr}. In
the special case of asymptotically flat algebraically special metrics, \{or
in their flat-space limits\} one could choose null infinity, $\mathfrak{I}%
^{+},$ as the realization space of the CR manifold where a `portion' of the
metric (coming from the congruence itself) defines the CR structure.

Our complex world-lines arise from the special class of shear-free null
geodesic congruences that are `regular' in the neighborhood of future null
infinity, i.e., $\mathfrak{I}^{+}.$ By `regular' we mean that all the rays
pierce $\mathfrak{I}^{+}$ and are never tangent to it. Specifically, we have
the result\cite{CR,footprints} that every regular shear-free null geodesic
congruence in Minkowski space is generated by a complex world-line in
complex Minkowski space and every complex world-line determines a CR
function and structure.

We show that for every asymptotically flat Maxwell field there is a unique
regular (at $\mathfrak{I}^{+}$) shear-free null geodesic congruence whose
tangent vectors determine a null tetrad system at $\mathfrak{I}^{+}$ such
that the tetrad components of the Maxwell field have a vanishing electric
and magnetic dipole moment. A one-to-one relationship between the complex
world-line and the original pair of dipoles is given.

Since the material discussed here is far from the mainstream, section II
will be devoted to an extensive preliminary discussion of null infinity,
shear free null geodesic congruences and their relationship to CR
structures. In section III we show how the \textit{complex} world-line can
be decomposed into a \textit{real} world-line with further \textit{real}
structure. In section IV the relationship between the Maxwell field and the
complex world-line is described. In the conclusion, section V, we discuss
how the ideas described here can be transferred to asymptotically flat
Einstein and Einstein-Maxwell metrics.

\section{Preliminaries}

\subsection{Null Infinity}

We begin with a brief review of Penrose's null infinity, $\mathfrak{I}^{+},$
the endpoints of all future directed null geodesics for either Minkowski
space or for asymptotically flat space-times. Though it was obtained,
formally, by the conformal compactification of the physical space-time, the
intuitive picture of the `space' of the future infinity limit of the affine
length of the null geodesics gives an excellent picture for our purposes.

$\mathfrak{I}^{+}$ is a null three-surface, $S^{2}xR$, that is naturally
coordinatized by the `so-called' Bondi coordinates\cite{Bondi}, $(u,\zeta ,%
\overline{\zeta })$ where ($\zeta ,\overline{\zeta }$) are complex
stereographic coordinates on the sphere, $S^{2},$ that label the null
generators of $\mathfrak{I}^{+},$ while $u$ labels the uniformly separated
cuts of $\mathfrak{I}^{+}.$ The freedom in the choice of Bondi coordinates
is the so-called BMS group. Though we will use an arbitrary but fixed Bondi
coordinate system and not concern ourselves with BMS transformations, our
results are independent of the choice of Bondi frame.

With the Bondi coordinates there is a naturally associated null tetrad
system, $(l^{a},n^{a},m^{a},\overline{m}^{a})$: where $l^{a}$ is the past
null direction normal to the $u$ constant cut, at each point of $\mathfrak{I}%
^{+},$ $n^{a}$ is the tangent vector to the generators of $\mathfrak{I}^{+}$%
, while $(m^{a},\overline{m}^{a})$ are the (complex) tangent vectors to the
cuts, $u=const.$

Often one find use for a different coordinatization of $\mathfrak{I}^{+},$
referred to as NU coordinates, where the generators keep the coordinates ($%
\zeta ,\overline{\zeta })$ but the slicing is given by the relation,
\begin{equation}
u=G(s,\zeta ,\overline{\zeta })  \label{NU}
\end{equation}
where constant $s$'s yield arbitrarily spaced, $S^{2},$ slices. An important
special case of Eq.(\ref{NU}) are the cuts obtained from the one-parameter
family of future directed light-cones emanating from a given time-like
world-line, say e.g., $x^{a}=\xi ^{a}(s).$

For each point on $\mathfrak{I}^{+}$ there is the past sphere of all null
directions which, aside from the one direction tangent to $\mathfrak{I}^{+},$
point `backwards' into the space-time. Since they form the sphere of null
directions, they can be labeled by their own stereographic coordinates which
we will denote by $L.$ If we have a field of past null directions, they can
be described by the `field' of stereographic angles, $L$($u,\zeta ,\overline{%
\zeta }$). The direction $L(u,\zeta ,\overline{\zeta })=\infty $ corresponds
to the direction $n^{a}$ \ while $L=0$ corresponds to the direction of the
Bondi ray, $l^{a}$.

\subsection{Null Geodesic Congruences}

Since one of our major items for investigation are null geodesic congruences
in Minkowski space, we begin by giving an analytic description of any
arbitrary null geodesic congruence, (with $l^{a}$ as tangent vector)\cite%
{Sh.Free.Max} . Using $x^{a}$ as standard Minkowski-space coordinates, $%
L(u,\zeta ,\overline{\zeta })$ and $r_{0}(u,\zeta ,\overline{\zeta })$ as
arbitrary complex functions of the `parameters' ($u,\zeta ,\overline{\zeta }$%
) which label individual members of the null geodesic congruence and $r$ as
the affine parameter along each null geodesic, we find that any null
geodesic congruence can be described by
\begin{eqnarray}
x^{a} &=&u(l^{a}+n^{a})-L\overline{m}^{a}-\overline{L}m^{a}+(r-r_{0})l^{a},
\label{ngc''} \\
u_{r} &=&\sqrt{2}u,
\end{eqnarray}%
with $u_{r}$ the retarded time, $(t-r),$ $u$ the Bondi time and the ($\zeta ,%
\overline{\zeta }$) dependent null tetrad given by
\begin{eqnarray}
l^{a} &=&\frac{\sqrt{2}}{2P}(1+\zeta \overline{\zeta },\text{ }\zeta +%
\overline{\zeta },\text{ }-i(\zeta -\overline{\zeta }),-1+\zeta \overline{%
\zeta });  \label{tetrad} \\
m^{a} &=&\text{\dh }l^{a}=\frac{\sqrt{2}}{2P}(0,1-\overline{\zeta }^{2},-i(1+%
\overline{\zeta }^{2}),\text{ }2\overline{\zeta }),  \nonumber \\
\overline{m}^{a} &=&\overline{\text{\dh }}l^{a}=\frac{\sqrt{2}}{2P}%
(0,1-\zeta ^{2},\text{ }i(1+\zeta ^{2}),2\zeta ),  \nonumber \\
n^{a} &=&\frac{\sqrt{2}}{2P}(1+\zeta \overline{\zeta },-(\zeta +\overline{%
\zeta }),\text{ }i(\zeta -\overline{\zeta }),1-\zeta \overline{\zeta }),
\nonumber \\
t^{a} &=&\frac{l^{a}+n^{a}}{\sqrt{2}},  \label{t} \\
P &=&1+\zeta \overline{\zeta }.  \label{P}
\end{eqnarray}

The optical parameters\cite{NP}, i.e., the complex divergence $\rho ,$ the
complex shear $\sigma \ $and the twist $\Sigma ,$ are given by
\begin{eqnarray}
\rho &=&\frac{i\Sigma -r}{r^{2}+\Sigma ^{2}-\overline{\sigma }_{0}\sigma _{0}%
}  \label{optical} \\
\sigma &=&\frac{\sigma _{0}}{r^{2}+\Sigma ^{2}-\overline{\sigma }_{0}\sigma
_{0}}  \nonumber
\end{eqnarray}%
with
\begin{eqnarray}
\sigma _{0} &=&\text{\dh }L+LL,_{u}  \label{sigmao} \\
2i\Sigma &=&\text{\dh }\overline{L}+L\overline{L},_{u}-\text{ }\overline{%
\text{\dh }}L-\overline{L}L,_{u}\text{ .}  \label{SIGMA}
\end{eqnarray}%
The function $r_{0}\ $has been chosen as
\[
r_{0}=-\frac{1}{2}(\text{\dh }\overline{L}+L\overline{L},_{u}+\overline{%
\text{\dh }}L+\overline{L}L,_{u})
\]%
in order to simplify~$\rho $ by choosing an origin for the affine parametter
$r$.

We thus see that the optical parameters are determined by the choice of $%
L(u,\zeta ,\overline{\zeta }).$

Our main interest lies in the class of \textit{regular} null geodesic
congruences with a vanishing shear. This condition is achieved by first
imposing the shear-free condition on $L,$ i.e., that $L$ satisfy the
differential condition\cite{A},
\begin{equation}
\text{\dh }L+LL,_{u}=0  \label{Shearfree}
\end{equation}
which leads immediately to
\begin{equation}
\rho =\frac{i\Sigma -r}{r^{2}+\Sigma ^{2}}=-\frac{1}{r+i\Sigma }.
\label{rho}
\end{equation}

There are two immediate remarks to be made:\\

{\bf Remark 1.} From Eq.(\ref{rho}) we see that the caustic set of
the congruence is determined \noindent \noindent by the conditions,
$r=0$ and $\Sigma (u,\zeta ,\overline{\zeta })=0.$These two
conditions define two three-surfaces in Minkowski space. If these
conditions are put into Eq.(\ref{ngc''}) we have the explicit
parametric description of the caustic set in terms of the Minkowski
$x^{a}.$\\

{\bf Remark 2.} The second remark is of the utmost importance. We
stress that the complex function $L(u,\zeta ,\overline{\zeta })$
used to determine the null geodesic congruence becomes, in the limit
of $r\rightarrow \infty ,$ i.e., as the geodesics approach
$\mathfrak{I}^{+},$ the complex stereographic field
described in the previous subsection and the geodesic parameters $(u,\zeta ,%
\overline{\zeta })$ become the Bondi coordinates of
$\mathfrak{I}^{+}.$\\

\subsection{Regular Shear-free Congruences}

Though any solution to Eq.(\ref{Shearfree}) determines a null geodesic
congruence, we consider only the `regular' ones. `Regular' means an $%
L(u,\zeta ,\overline{\zeta })$ that has no infinities, i.e. the $L(u,\zeta ,
\overline{\zeta })$ does not contain any directions that are tangent to $%
\mathfrak{I}^{+}.$ The regular congruences are obtained in the following
manner:

We introduce the complex function
\begin{equation}
\tau =T(u,\zeta ,\overline{\zeta })  \label{phi}
\end{equation}
via solutions to
\begin{equation}
\text{\dh }T+L\,T,_{{\small u}}=0.  \label{Lofphi}
\end{equation}
($T$ is a CR function which, with $\widetilde{\zeta }=$ $\overline{\zeta },$
endows $\mathfrak{I}^{+}$ with a CR structure. See the appendix for a brief
discussion of CR structures.)

Direct substitution of
\begin{equation}
L\,(u,\zeta ,\overline{\zeta })=-\frac{\text{\dh }T}{T,_{{\small u}}}
\label{L1}
\end{equation}%
into Eq.(\ref{Shearfree}) does not simplify it at all.\textit{\ }However if
we analytically extend $u$ into the complex domain in Eq.(\ref{phi}) and
reverse the point of view and formally treat $\tau $ as the \textit{%
independent variable} instead of $\ u,$ so that when Eq.(\ref{phi}) is
rewritten \textit{implicitly} as

\begin{equation}
u=X(\tau ,\zeta ,\overline{\zeta }),  \label{U}
\end{equation}%
a huge simplification occurs.

{\bf Remark 3.} It must be emphasized that though we are concerned
with real values of $u,$ due to the assumed analyticity of our
expressions, Eq.(\ref{U} ) allows for
complex values of $u.$ Later in this section we will find those values of $%
\tau $ that yield real values of $u.$\\

Differentiating Eq.(\ref{U}) we have
\begin{eqnarray}
1 &=&X,_{\tau }(\tau ,\zeta ,\overline{\zeta })\,T,_{{\small u}},
\label{XDerivates} \\
0 &=&X_{\tau }(\tau ,\zeta ,\overline{\zeta })\text{\dh }T+\text{\dh }%
_{(\tau )}X,  \nonumber
\end{eqnarray}%
\{\dh $_{(\tau )}$ means the edth derivative holding $\tau $ constant rather
than holding $u$ constant\} so that
\begin{equation}
L\,=-\frac{\text{\dh }T}{T,_{{\small u}}}=\text{\dh }_{(\tau )}X.
\label{LofX}
\end{equation}

Using Eqs.(\ref{LofX}) and (\ref{XDerivates}), Eq.(\ref{Shearfree}) becomes
the easily integrable equation

\begin{equation}
\text{\dh }_{(\tau )}^{2}X=0  \label{edth2X}
\end{equation}
whose general regular solution has the form

\begin{equation}
u=X(\tau ,\zeta ,\overline{\zeta })=\frac{\alpha (\tau )+\beta (\tau )%
\overline{\zeta }+\overline{\beta }(\tau )\zeta +\gamma (\tau )\zeta
\overline{\zeta }}{1+\zeta \overline{\zeta }}.  \label{non-sing}
\end{equation}

In terms of a spherical harmonic expansion of $X(\tau ,\zeta ,\overline{%
\zeta }),$ this expression involves just the $(l=0,1)$ terms with the four
coefficients given as arbitrary complex analytic functions of $\tau .$ Note
that the four components of $l_{a}(\zeta ,\overline{\zeta })$ are linear
combinations of the four $(l=0,1)$ harmonics which leads to the concise (and
useful) expression,
\begin{equation}
u=X(\tau ,\zeta ,\overline{\zeta })=\xi ^{a}(\tau )l_{a}(\zeta ,\overline{%
\zeta })  \label{sol2}
\end{equation}
which implicitly determines
\begin{equation}
\tau =T(u,\zeta ,\overline{\zeta }).  \label{T*}
\end{equation}

From this and Eq. (\ref{LofX}), we see that the regular solutions to the
Eq.( \ref{Shearfree}) are given in the parametric form
\begin{eqnarray}
u &=&X(\tau ,\zeta ,\overline{\zeta })\equiv \xi ^{a}(\tau )l_{a}(\zeta ,%
\overline{\zeta }),  \label{L&u} \\
L(u,\zeta ,\overline{\zeta }) &=&\text{\dh }_{(\tau )}X(\tau ,\zeta ,%
\overline{\zeta })=\xi ^{a}(\tau )m_{a}(\zeta ,\overline{\zeta }).  \nonumber
\end{eqnarray}

\{One must be certain that $V\equiv \partial _{\tau }X=\partial _{\tau }\xi
^{a}(\tau )l_{a}(\zeta ,\overline{\zeta })=$ $v^{a}(\tau )l_{a}(\zeta ,
\overline{\zeta })$ is different from zero for those values of $\tau $ that
yield real values of $u$.\}

Aside from this weak condition we have: An arbitrary complex world-line in
complex Minkowski space determines a shear-free null geodesic congruence
that is regular at $\mathfrak{I}^{+}.$

For later reference we also compute the twist since it adopts a very simple
form when expressed in terms of $X$ and $\overline{X}$.

Starting from Eq.(\ref{SIGMA}),
\[
2i\Sigma =\text{\dh }_{(u)}\overline{L}+L\overline{L},_{u}-\text{ }\overline{%
\text{\dh }}_{(u)}L-\overline{L}L,_{u}\text{ .}
\]%
and using the following useful identity
\[
\overline{\text{\dh }}_{(\tau ,\overline{\tau })}\text{\dh }_{(\tau ,%
\overline{\tau })}X=\overline{\text{\dh }}_{\tau }L=\overline{\text{\dh }}%
_{u}L+L,_{u}\overline{\text{\dh }}_{(\tau )}\overline{X}=\overline{\text{\dh
}}_{(u)}L+\overline{L}L,_{u},
\]%
and its c.c., one can write $\Sigma $ as%
\begin{equation}
2i\Sigma =\overline{\text{\dh }}_{(\tau ,\overline{\tau })}\text{\dh }%
_{(\tau ,\overline{\tau })}(X-\overline{X})=(\xi ^{a}(\tau )-\overline{\xi }%
^{a}(\overline{\tau }))\left( n_{a}-l_{a}\right) ,  \label{twist*}
\end{equation}

{\bf Remark 4.} For any given $\tau =T(u,\zeta ,\overline{\zeta }),$
we see that
\begin{equation}
\tau ^{\ast }=\Phi (\tau ),  \label{gaugefreedom}
\end{equation}

with arbitrary analytic $\Phi (\tau ),$ yields the same $L,$ since
\begin{eqnarray}
\text{\dh }\tau ^{\ast } &=&\partial _{\tau }\Phi \cdot \text{\dh }\tau
\label{reparametrize} \\
\tau ^{\ast },_{{\small u}} &=&\partial _{\tau }\Phi \,\cdot \,\tau ,_{%
{\small u}}  \nonumber
\end{eqnarray}%
and
\[
L=-\frac{\text{\dh }T^{\ast }}{T^{\ast },_{{\small u}}}=-\frac{\text{\dh }T}{%
T,_{{\small u}}}.
\]

This freedom in the choice of $\tau ,$ which will be very useful
later, is referred to as the `regauging' of $\tau $. This is a
well-known feature of CR functions. Though we \ could have locally
introduced $\Phi $ as an arbitrary function of both $\tau $ $and$
$\overline{\zeta },$ this would have introduced angular
singularities.\\

{\bf Remark 5.} When the world-line is given as a real function of a
real variable, then the twist vanishes, i.e., $\Sigma =0,$ and the
caustic set is simply the real world-line$.$\\

{\bf Remark 6.} It should be emphasized that much of the material of
this section is not new. It has been developed by many workers and
has appeared in many versions. For example, finding solutions to
Eq.(\ref{Shearfree}) was a special case of an early result in
Penrose's Twistor theory\cite{RP3} and is often referred to as a
special case of the Kerr theorem.\\

\section{Real Structures from the CR Structure}

\subsection{Exact Real Informtion}

In this section we will extract real information from the complex CR
function $\tau =T(u,\zeta ,\overline{\zeta }).$ More specifically we will
obtain a real Minkowski world-line with a certain real structure attached to
the line. Later we will see that this attached structure can carry physical
information.

The difficulty with this extraction process lies in the gauge freedom, Eq.(%
\ref{gaugefreedom}),
\begin{equation}
\tau =>\tau ^{\ast }=\Phi (\tau )  \label{gauge}
\end{equation}%
where $\Phi (\tau )$ is an \textit{arbitrary} complex analytic function. We
wish to find a canonical choice for $\tau $ and thus eliminate this gauge
freedom$.$ We first note that if the Taylor series for a function of $\tau $
has complex (or real) coefficients we will refer to it as a complex (or
real) analytic function. To find the canonical choice we take advantage of a
further structure that is available, namely the Minkowski space norm that is
applied to the complex velocity vector of the world-line, i.e., to
\[
v^{a}=\xi ^{a},_{\tau }(\tau )\equiv \xi ^{a\,\prime }(\tau ).
\]%
The argument requires two stages. Using the gauge freedom, Eq. (\ref{gauge}%
), the velocity vector transforms as
\[
v^{\ast \,a}=\xi ^{a},_{\tau \ast \,}(\tau ^{\ast }\,)=[\Phi (\tau )^{\prime
}]^{-1}\xi ^{a\prime }=[\Phi (\tau )^{\prime }]^{-1}v^{a},
\]%
so that the Minkowski norm $v^{2}=\eta _{ab}v^{a}v^{b}$ transforms as
\[
v^{\ast \,2}=[\Phi (\tau )^{\prime }]^{-2}v^{2}.
\]%
By choosing
\begin{equation}
\Phi (\tau )=\int^{\tau }\sqrt{v^{2}(\tau ^{\prime })}d\tau ^{\prime }
\label{PHI}
\end{equation}%
we have a complex $v^{\ast \,a}$ with a unit norm. The next stage, after
dropping the ($^{\ast }$), is to write $v^{\,a}(\tau )=v_{R}^{\,a}(\tau
)+iv_{I}^{\,a}(\tau )$ where both $v_{R}^{\,a}(\tau )$ and $v_{I}^{\,a}(\tau
)$ are \textit{real analytic} vectors, i.e. both have Taylor series with
real coefficients and are orthogonal to each other. The purpose of
introducing the gauging with Eq.(\ref{PHI}) was to allow the decomposition
of $v^{\,a}(\tau )$ into its orthogonal `real' and `imaginary' parts. Any
further \textit{real analytic} regauging preserves this decomposition.
Finally, by performing a new gauge transformation $\tau ^{\ast }=\Theta
(\tau ),$ with the function $\Theta (\tau )$ restricted to \textit{real
analytic}, we introduce our canonical choice of $\tau $ (again dropping the
*) by choosing the `real' velocity vector, $v_{R}^{\,a}(\tau ),$ to have a
unit norm, $\eta _{ab}v_{R}^{\,a}(\tau )v_{R}^{\,\,b}(\tau )=1$. $\tau $ is
thus uniquely determined up to the choice of an additive constant.

{\bf Remark 7.} It is easy to see from the fact that now $\eta
_{ab}v^{\,a}(\tau )v^{\,\,b}(\tau )$ though no longer equal to one,
is still a real analytic function of $\tau $ from which it follows
that $\eta _{ab}v_{R}^{\,a}(\tau )v_{I}^{\,b}(\tau )=0.$\\

Using this canonical choice of $\tau $ we write it as $\tau =s+i\lambda $
and decompose our complex functions into their real and imaginary parts.
Specifically we write the CR function $T(u,\zeta ,\overline{\zeta })$ and
inverse $X(\tau ,\zeta ,\overline{\zeta })$ as
\begin{eqnarray}
\tau &=&s+i\lambda =T(u,\zeta ,\overline{\zeta })=T_{R}(u,\zeta ,\overline{%
\zeta })+iT_{I}(u,\zeta ,\overline{\zeta })  \label{T_RI} \\
u &=&X(\tau ,\zeta ,\overline{\zeta })=X_{R}(\tau ,\zeta ,\overline{\zeta }%
)+iX_{I}(\tau ,\zeta ,\overline{\zeta })  \label{X_RI} \\
&=&X_{R}(s+i\lambda ,\zeta ,\overline{\zeta })+iX_{I}(s+i\lambda ,\zeta ,%
\overline{\zeta })  \nonumber
\end{eqnarray}
where $T_{R},$ $T_{I},$ $X_{R}$ and $X_{I}$ are all real analytic functions,
$s$ and $\lambda $ are real variables and ($u,\zeta ,\overline{\zeta }$) are
the `real' coordinates of real $\mathfrak{I}^{+}.$ One has immediately, from
Eq. (\ref{T_RI}), that
\begin{eqnarray*}
s &=&T_{R}(u,\zeta ,\overline{\zeta }) \\
\lambda &=&T_{I}(u,\zeta ,\overline{\zeta }).
\end{eqnarray*}
However, since we know the detailed form $X(\tau ,\zeta ,\overline{\zeta }
)=\xi ^{a}(\tau )l_{a}(\zeta ,\overline{\zeta }),$ it is more useful to work
with Eq.(\ref{X_RI}). We decompose $\xi ^{a}(\tau )$ as
\[
\xi ^{a}(\tau )=\xi _{R}^{a}(\tau )+i\xi _{I}^{a}(\tau )
\]
or
\[
\xi ^{a}(\tau )=\xi _{R}^{a}(s+i\lambda )+i\xi _{I}^{a}(s+i\lambda ).
\]

Our goal is to write $\xi ^{a}(\tau )$ as
\begin{eqnarray}
\xi ^{a}(\tau ) &=&\chi ^{a}(s,\lambda )=\chi _{R}^{a}(s,\lambda )+i\chi
_{I}^{a}(s,\lambda )  \label{xi} \\
&=&\sum \frac{1}{(2n)!}f_{2n}(-1)^{n}\lambda ^{2n}+i\sum \frac{1}{(2n+1)!}%
f_{2n+1}(-1)^{n}\lambda ^{2n+1}  \nonumber
\end{eqnarray}%
where $\chi _{R}^{a}(s,\lambda )$ and $\chi _{I}^{a}(s,\lambda )$ are real
functions. This is easily accomplished by expanding both $\xi
_{R}^{a}(s+i\lambda )$ and $\xi _{I}^{a}(s+i\lambda )$ in a Taylor series in
$i\lambda $ and regrouping the terms, i.e.,

\begin{eqnarray}
\xi _{R}^{a}(s+i\lambda ) &=&\Sigma _{n}\frac{(-1)^{n}}{(2n)!}\xi
_{R}^{a\,(2n)}(s)\lambda ^{2n}+i\Sigma _{n}\frac{(-1)^{n}}{(2n+1)!}\xi
_{R}^{a\,(2n+1)}(s)\lambda ^{2n+1}  \label{taylor series} \\
i\xi _{I}^{a}(s+i\lambda ) &=&i\Sigma _{n}\frac{(-1)^{n}}{(2n)!}\xi
_{I}^{a\,(2n)}(s)\lambda ^{2n}-\Sigma _{n}\frac{(-1)^{n}}{(2n+1)!}\xi
_{I}^{a\,(2n+1)}(s)\lambda ^{2n+1}  \nonumber \\
\xi ^{a}(\tau ) &=&\Sigma _{n}[\frac{(-1)^{n}}{(2n)!}\xi
_{R}^{a\,(2n)}(s)\lambda ^{2n}-\frac{(-1)^{n}}{(2n+1)!}\xi
_{I}^{a\,(2n+1)}(s)\lambda ^{2n+1} \\
&&+i\{\frac{(-1)^{n}}{(2n+1)!}\xi _{R}^{a\,(2n+1)}(s)\lambda ^{2n+1}+\frac{%
(-1)^{n}}{(2n)!}\xi _{I}^{a\,(2n)}(s)\lambda ^{2n}]  \nonumber
\end{eqnarray}%
so that we have
\begin{eqnarray}
\xi ^{a}(\tau ) &=&\chi _{R}^{a}(s,\lambda )+i\chi _{I}^{a}(s,\lambda )
\label{xi*} \\
\chi _{R}^{a}(s,\lambda ) &=&\Sigma _{n}[\frac{(-1)^{n}}{(2n)!}\xi
_{R}^{a\,(2n)}(s)\lambda ^{2n}-\frac{(-1)^{n}}{(2n+1)!}\xi
_{I}^{a\,(2n+1)}(s)\lambda ^{2n+1}]  \label{xiR} \\
\chi _{I}^{a}(s,\lambda ) &=&\Sigma _{n}[\frac{(-1)^{n}}{(2n+1)!}\xi
_{R}^{a\,(2n+1)}(s)\lambda ^{2n+1}+\frac{(-1)^{n}}{(2n)!}\xi
_{I}^{a\,(2n)}(s)\lambda ^{2n}].  \label{xiI}
\end{eqnarray}%
and
\begin{eqnarray}
u &=&X(\tau ,\zeta ,\overline{\zeta })=\xi ^{a}(\tau )l_{a}(\zeta ,\overline{%
\zeta })  \label{Xdecomposed} \\
&=&\chi _{R}^{a}(s,\lambda )l_{a}(\zeta ,\overline{\zeta })+i\chi
_{I}^{a}(s,\lambda )l_{a}(\zeta ,\overline{\zeta })  \nonumber \\
&=&\chi _{R}(s,\lambda ,\zeta ,\overline{\zeta })+i\chi _{I}(s,\lambda
,\zeta ,\overline{\zeta }).  \nonumber
\end{eqnarray}%
For real values of $u$ we have the implicit equation for the determination
of $\lambda ,$ i.e.,
\begin{equation}
\chi _{I}(s,\lambda ,\zeta ,\overline{\zeta })=0  \label{xiI=0}
\end{equation}%
or solving for $\lambda ,$
\begin{equation}
\lambda =\Lambda (s,\zeta ,\overline{\zeta }).  \label{LAMBDA}
\end{equation}

\begin{description}
\item[Definition] In order to avoid the ambiguity of whether $u$ is real or
complex, in the following we will write $u_{C}$ for the complex $u$.
\end{description}

Finally we see that the information contained in the normalized complex
world-line, $\xi ^{a}(\tau ),$ and the associated complex cuts, $u_{C}=\xi
^{a}(\tau )l_{a}(\zeta ,\overline{\zeta }),$ is equivalently given by the
family of real-cuts of $\mathfrak{I}^{+},$
\begin{equation}
u=\chi _{R}(s,\Lambda (s,\zeta ,\overline{\zeta }),\zeta ,\overline{\zeta }%
)=\chi (s,\zeta ,\overline{\zeta })  \label{realu}
\end{equation}%
with the additional real structure $\lambda =\Lambda (s,\zeta ,\overline{%
\zeta })$ which carries the `imaginary' information that is `hidden' in the
complex world-line.

In order to find the twisting null congruence, (or rather the angle field $%
L(u,\zeta ,\overline{\zeta })$) in this real version, we must apply the \dh\ %
operator, holding both $(s,\lambda )$ constant, first to Eq.(\ref%
{Xdecomposed}), i.e. to $u_{C},$ and then eliminate the $\lambda $ via Eq.(%
\ref{LAMBDA}). This yields $L$ as a function of $(s,\zeta ,\overline{\zeta }%
).$

{\bf Remark 8.} If the \dh\ operator is first applied to
Eq.(\ref{realu}) it leads to a null geodesic congruence that is
\textit{not, in general, }shear-free but is instead twist-free. In
other words it would be incorrect to use
\[
L^{\ast }(s,\zeta ,\overline{\zeta })=\text{\dh }_{(s)}\chi _{R}(s,\Lambda
,\zeta ,\overline{\zeta })
\]%
for the twisting shear-free angle field. One must differentiate
first before setting $\chi _{I}=0.$ On the other hand $L^{\ast
}(s,\zeta ,\overline{\zeta })$ is in its own right an interesting
angle field.\\

The stereographic angle field, $L(u,\zeta ,\overline{\zeta }),$ is now given
in terms of ($s,\zeta ,\overline{\zeta })$ by
\begin{eqnarray}
L(s,\zeta ,\overline{\zeta }) &=&\text{\dh }_{(s,\lambda )}\chi
_{R}(s,\lambda ,\zeta ,\overline{\zeta })+i\text{\dh }_{(s,\lambda )}\chi
_{I}(s,\lambda ,\zeta ,\overline{\zeta }),  \label{L(s,zeta)} \\
\lambda &=&\Lambda (s,\zeta ,\overline{\zeta }).  \nonumber
\end{eqnarray}%
If wanted one could find $L$ as a function of $(u,\zeta ,\overline{\zeta })$
by eliminating the $s$ in Eq.(\ref{L(s,zeta)}) via the inverse of Eq.(\ref%
{realu}).

\subsection{Approximating the real information}

We can obtain a better understanding of the results of the previous
subsection by considering $\lambda $ to be small and look at only the first
order terms in $\lambda $ in the Taylor series. This leads to
\begin{eqnarray}
\xi ^{a}(\tau ) &=&\chi _{R}^{a}(s,\lambda )+i\chi _{I}^{a}(s,\lambda )
\label{1st order} \\
\chi _{R}^{a}(s,\lambda ) &=&\xi _{R}^{a\,}(s)-\xi _{I}^{a\,\prime
}(s)\lambda  \label{R} \\
\chi _{I}^{a}(s,\lambda ) &=&\xi _{I}^{a\,}(s)+\xi _{R}^{a\,\prime
}(s)\lambda  \label{I}
\end{eqnarray}%
or

\begin{equation}
u_{C}=[\xi _{R}^{a\,}(s)-\xi _{I}^{a\,\prime }(s)\lambda ]l_{a}+i[\xi
_{I}^{a\,}(s)+\xi _{R}^{a\,\prime }(s)\lambda ]l_{a}.  \label{Xdecomposed*}
\end{equation}%
Eq.(\ref{xiI=0}) implies that
\[
\lbrack \xi _{I}^{a\,}(s)+\xi _{R}^{a\,\prime }(s)\lambda ]l_{a}=0
\]%
or
\begin{equation}
\lambda (s,\zeta ,\overline{\zeta })=-\frac{\xi _{I}^{b\,}(s)l_{b}}{\xi
_{R}^{c\,\prime }(s)l_{c}},  \label{LAMBDA*}
\end{equation}%
so the real $u,$ Eq.(\ref{realu}), to linear order is
\begin{equation}
u=\xi _{R}^{a\,}(s)l_{a}+\xi _{I}^{a\,\prime }(s)l_{a}\frac{\xi
_{I}^{b\,}(s)l_{b}}{\xi _{R}^{c\,\prime }(s)l_{c}}  \label{realu*}
\end{equation}%
with
\begin{equation}
V=[\xi _{R}^{a\,\prime }(s)+\xi _{I}^{a\,\prime \prime }(s)\frac{\xi
_{I}^{b\,}(s)l_{b}}{\xi _{R}^{c\,\prime }(s)l_{c}}]l_{a}+i[\xi
_{I}^{a\,\prime }(s)-\xi _{R}^{a\,\prime \prime }(s)\frac{\xi
_{I}^{b\,}(s)l_{b}}{\xi _{R}^{c\,\prime }(s)l_{c}}]l_{a}.  \label{approxV}
\end{equation}

We see Eq.(\ref{realu*}) is an example of an NU slicing that was referred to
earlier. Though it is close to the special NU slicing that arises from the
light-cones emanating from a real world, $x^{a}=\xi _{R}^{a\,}(s),$ the
second term is the obstruction. The stereographic `angle', $L,$ as a
function of ($s,\zeta ,\overline{\zeta }$), is found by applying \dh\ to Eq.(%
\ref{Xdecomposed*}) and eliminating $\lambda $ via Eq.(\ref{LAMBDA*})
yielding
\begin{equation}
L(s,\zeta ,\overline{\zeta })=\{\xi _{R}^{a}(s)+\frac{\xi _{I}^{b}(s)l_{b}}{%
v_{R}^{c}(s)l_{c}}v_{I}^{a}(s)+i[\xi _{I}^{a}(s)-\frac{\xi _{I}^{b}(s)l_{b}}{%
v_{R}^{c}(s)l_{c}}v_{R}^{a}(s)]\}m_{a}.  \label{approxL}
\end{equation}%
We have used for the real unit Lorentzian velocity vector
\[
\xi _{R}^{c\prime }(s)=v_{R}^{c}(s).
\]%
Note that in the rest frame since $v_{R}^{c}(s)=(1,0,0,0),$ we obtain
\[
L(s,\zeta ,\overline{\zeta })=\{\xi _{R}^{a}(s)+i\xi _{I}^{a}(s)+\sqrt{2}\xi
_{I}^{b}(s)l_{b}v_{I}^{a}(s)\}m_{a}.
\]

{\bf Remark 9.} In the approximation expressions, Eqs.(\ref{LAMBDA*}), (\ref{realu*}) and (%
\ref{approxL}) we see that both the positions and velocities of the
world-line appear at each value of $s.$ If higher approximations were used,
higher derivatives of the world-line would appear. In other words the full
information of the world-line would be contained in these expressions.
\\

\bigskip It is now a straightforward calculation to evaluate, in this linear
approximation, the twist, i.e., Eq.(\ref{SIGMA}) or Eq.(\ref{twist*}). Up to
order $\lambda $, the expression for $\Sigma $\ is given by

\begin{equation}
\Sigma (s,\zeta ,\overline{\zeta })=(\xi _{I}^{a}(s)+\lambda
(s)v_{R}^{a}(s))\left( n_{a}-l_{a}\right) =(\xi _{I}^{a}(s)-\frac{\xi
_{I}^{b\,}(s)l_{b}}{v_{R}^{c}(s)l_{c}}v_{R}^{a}(s))\left( n_{a}-l_{a}\right)
.  \label{twist}
\end{equation}%
In the rest frame this is an expression involving only the $l=1$ spherical
harmonics with $s$ dependent coefficients. The vanishing of $\Sigma $ is
thus an evolving $s$ dependent topological circle.

We thus have the theorem that the caustic set for any regular shear-free
null geodesic congruence in Minkowski space that is sufficiently close to
the shear and twist-free case is $SxR,$ i.e., a deformed circle moving in
time. This is simply a deformation of the circular caustic set of the
flat-space limit of the charged Kerr metric.

\section{Applications to Maxwell Fields}

\subsection{Bondi Form of Maxwell's Equations}

Before describing the applications of the shear-free congruences to Maxwell
theory, we first review some elementary facts of Maxwell fields transformed
to the spin-coefficient notation\cite{TodNew}.

The Maxwell tensor is given in the spin-coefficient notation by the three
complex tetrad components

\begin{eqnarray}
\phi _{0} &=&F_{ab}l^{a}m^{b}  \label{MaxField} \\
\phi _{1} &=&\frac{1}{2}F_{ab}(l^{a}n^{b}+m^{a}\overline{m}^{b})  \nonumber
\\
\phi _{2} &=&F_{ab}\overline{m}^{a}n^{b}.  \nonumber
\end{eqnarray}

The Bondi tetrad system $(l^{a},n^{a},m^{a},\overline{m}^{a})$ will, in the
remainder, be denoted by $(l_{B}^{a},n_{B}^{a},m_{B}^{a},\overline{m}
_{B}^{a}).$

The Maxwell equations, written in a Bondi tetrad frame with Bondi
coordinates, yield the asymptotic behavior
\begin{eqnarray}
\phi _{0B} &=&\frac{\phi _{0B}^{0}}{r^{3}}+O(r^{-4})  \label{BondiMax} \\
\phi _{1B} &=&\frac{\phi _{1B}^{0}}{r^{2}}+O(r^{-3})  \nonumber \\
\phi _{2B} &=&\frac{\phi _{2B}^{0}}{r}+O(r^{-2}).  \nonumber
\end{eqnarray}%
The asymptotic components, $\phi _{0B}^{0},$ $\phi _{1B}^{0},$ $\phi
_{2B}^{0},$ all functions ($u,\zeta ,\overline{\zeta }$), satisfy the
evolution equations,
\begin{eqnarray}
\phi _{0B}^{0\;\cdot }+\text{\dh }\phi _{1B}^{0} &=&0\;,  \label{lastMaxEqs}
\\
\phi _{1B}^{0\;\cdot }+\text{\dh }\phi _{2B}^{0} &=&0.  \nonumber
\end{eqnarray}%
Though given arbitrary (characteristic data) $\phi _{2B}^{0}(u,\zeta ,%
\overline{\zeta })$ these equations can be easily integrated, we will be
interested only in the monopole and dipole (electric and magnetic) terms,
i.e., just the $l=0,1$ harmonic components. This leads to the solution\cite%
{KN}, with an arbitrary \textit{complex} \textit{three-vector},
\begin{equation}
\hat{Z}^{i}(u)=\hat{Z}^{i}(\frac{u_{r}}{\sqrt{2}})\equiv Z^{i}(u_{r})
\label{Z}
\end{equation}

\begin{eqnarray}
\phi _{0B}^{0} &=&2q\hat{Z}^{i}Y_{1i}^{1}+H^{(1)}(l\geq 2)  \label{0} \\
\phi _{1B}^{0} &=&q+q(\hat{Z}^{i})^{\cdot }Y_{1i}^{0}+H^{(0)}(l\geq 2)
\label{1} \\
\phi _{2B}^{0} &=&-q(\hat{Z}^{i})^{\cdot \cdot }Y_{1i}^{-1}+H^{(-1)}(l\geq 2)
\label{2}
\end{eqnarray}

\begin{eqnarray*}
\hat{Z}^{i}(u)^{\cdot } &=&\sqrt{2}Z^{i}(u_{r})^{\prime } \\
\hat{Z}^{i}(u)^{\cdot \cdot } &=&2Z^{i}(u_{r})^{\prime \prime }
\end{eqnarray*}%
with $qZ^{i}(u_{r})$ being the complex (electric + $i$ magnetic) dipole
moments. The set ($Y_{1i}^{1},Y_{1i}^{0},Y_{1i}^{-1}$) are, respectively,
the $l=1$ parts of the spin $(1,0,-1)$ tensor harmonics\cite{spins}. The ($%
^{\cdot }$) and ($^{\prime }$) denote respectively the $u$ and $u_{r}$
derivatives.

{\bf Remark 10.} Often we encounter two functions that are
functionally identical but of different arguments, $u_{r}$ and s,
e.g., $Z^{a}(u_{r})$ and $Z^{a}(s).$ To denote differentiation we
will use prime ($^{\prime }$) for either variable.
\\

\subsection{Transformed Maxwell Field}

Under the following null rotation around $n_{B}^{a}$ (with $L(u,\zeta ,%
\overline{\zeta }),$ for the moment, an arbitrary angle field) at $\mathfrak{%
I}^{+}$

\begin{eqnarray}
l^{a} &=&l_{B}^{a}-\frac{\bar{L}}{r}m_{B}^{a}-\frac{L}{r}\overline{m}%
_{B}^{a}+\frac{L\bar{L}}{r^{2}}n_{B}^{a}+O(r^{-3}),  \label{NULLROT^} \\
m^{a} &=&m_{B}^{a}-\frac{L}{r}n_{B}^{a}+O(r^{-2}),  \nonumber \\
\overline{m}^{a} &=&\overline{m}_{B}^{a}-\frac{\bar{L}}{r}%
n_{B}^{a}+O(r^{-2}),  \nonumber \\
n^{a} &=&n_{B}^{a}  \label{B-T}
\end{eqnarray}%
the asymptotic components of the Maxwell field transform as

\begin{eqnarray}
\phi _{0}^{0} &=&\phi _{0B}^{0}-2L\;\phi _{1B}^{0}+L^{2}\;\phi _{2B}^{0},
\label{null rot 1} \\
\phi _{1}^{0} &=&\phi _{1B}^{0}-L\;\phi _{2B}^{0},  \label{null rot 2} \\
\phi _{2}^{0} &=&\phi _{2B}^{0}.  \label{null rot 3}
\end{eqnarray}
The asymptotic Maxwell equations, Eq.(\ref{lastMaxEqs}), become
\begin{eqnarray}
\phi _{0}^{0\cdot }+\text{\dh }\phi _{1}^{0}+2L^{\cdot }\phi _{1}^{0}+L\phi
_{1}^{0\cdot } &=&0,  \label{ME1T} \\
\phi _{1}^{0\cdot }+\text{\dh }\phi _{2}^{0}+(L\phi _{2}^{0})^{\cdot } &=&0.
\label{ME2T}
\end{eqnarray}

The angle field $L$ is then \textit{chosen} so that $l^{a}$ determines a
\textit{regular} shear-free null geodesic congruence, i.e., it has the form,
from Eqs.( \ref{L&u}),

\begin{eqnarray}
L(u,\zeta ,\overline{\zeta }) &=&\xi ^{a}(\tau )m_{a}(\zeta ,\overline{\zeta
})  \label{L&u*} \\
u &=&\xi ^{a}(\tau )l_{a}(\zeta ,\overline{\zeta }).  \nonumber
\end{eqnarray}%
The complex world-line is then determined from the Maxwell field in terms of
the functions $Z^{i}(u)$ by imposing the following condition:

$\blacktriangleright $We require the new $\phi _{0}^{0}$ to have \textit{%
vanishing (complex) dipole terms}, i.e., the $l=1$ harmonics of $\phi
_{0}^{0},$ in Eq.(\ref{null rot 1}), are required to vanish.

To \textit{exactly} implement this condition is very difficult, if possible,
needing many Clebsch-Gordon decompositions and the use of the Taylor series
for $\xi ^{a}(s+i\lambda ).$ To avoid this we will use several
approximations: we keep only the linear terms in $\lambda $ from the
previous section, a slow motion expansion (near the rest frame) for the
velocity vector of the real world-line $\xi _{R}^{a}(s)$ and finally we
treat all variables (except $q$) as small and keep only quadratic terms.
Since the velocity $v_{R}^{a}(s)=\xi _{R}^{a\,\prime }(s)$ has been
normalized to one, we can write
\begin{eqnarray}
v_{R}^{a}(s) &=&(v_{R}^{0}(s),v_{R}^{i}(s))  \nonumber \\
v_{R}^{0}(s) &=&\sqrt{1+(v_{R}^{i})^{2}}=1+\delta v_{R}^{0}\text{ }\approx
\text{ }1+\frac{1}{2}v^{i2}+..  \label{normalization}
\end{eqnarray}

\noindent and hence, within this approximation scheme, we have from Eq.(\ref%
{realu*}),
\begin{eqnarray*}
u &=&\xi _{R}^{a\,}(s)l_{a}+\xi _{I}^{a\,\prime }(s)l_{a}\frac{\xi
_{I}^{b\,}(s)l_{b}}{\xi _{R}^{c\,\prime }(s)l_{c}} \\
u &=&\xi _{R}^{a\,}(s)l_{a}+\sqrt{2}v_{I}^{a\,}(s)l_{a}\xi _{I}^{b\,}(s)l_{b}
\\
u_{r} &=&\sqrt{2}u=s-\frac{1}{\sqrt{2}}\xi
_{R}^{i}(s)Y_{1i}^{0}+2v_{I}^{a\,}(s)l_{a}\xi _{I}^{b\,}(s)l_{b}
\end{eqnarray*}

The complex three-vector, $Z^{i}(u_{r}),$ expressed as a function of $s,$
becomes
\[
Z^{j}(u_{r})=Z^{j}(s)-\frac{1}{\sqrt{2}}\xi _{R}^{i}(s)Y_{1i}^{0}Z^{j\prime
}(s).
\]%
The Eq.(\ref{null rot 1}),
\[
\phi _{0}^{0}=\phi _{0B}^{0}-2L\;\phi _{1B}^{0}+L^{2}\;\phi _{2B}^{0},
\]%
\textit{retaining only} the $l=1$ harmonics and neglecting the $L^{2}$ term,
$\phi _{0}^{0}[l=1]=0$ becomes

\begin{eqnarray}
0 &=&Z^{k}(s)-\xi _{R}^{k}(s)-i\xi _{I}^{k}(s)-\xi _{I}^{0}(s)\xi
_{I}^{k\prime }(s)+i\xi _{I}^{0}(s)\xi _{R}^{k\prime }(s)  \label{condition}
\\
&&-\epsilon _{ijk}{\large (}\frac{1}{2}i\xi _{R}^{i}(s)-\xi
_{I}^{i}(s)Z^{j\prime }(s){\large )}+\frac{1}{2}\epsilon _{ijk}{\large (}\xi
_{I}^{j}(s)\xi _{R}^{i\prime }(s)+i\xi _{I}^{j}(s)\xi _{I}^{i\prime }(s)%
{\large )}.  \nonumber
\end{eqnarray}%
This equation can be solved by a simple iteration process in either one of
two ways:

\textit{First assuming that the world-line }$\xi ^{k}(s)$ \textit{is known},
i.e., that
\begin{equation}
\xi ^{k}(s)=\xi _{R}^{k}(s)+i\xi _{I}^{k}(s)  \label{wordline s}
\end{equation}%
is given, then the linear solutions for $Z^{k}(s)$ is
\begin{equation}
Z^{k}(s)=\xi _{R}^{k}(s)+i\xi _{I}^{k}(s).  \label{linear sol}
\end{equation}%
This when substituted back into Eq.(\ref{condition}) leads to
\[
Z^{k}(s)=\xi _{R}^{k}(s)+i\xi _{I}^{k}(s)-i\xi _{I}^{0}\{\xi _{R}^{k\prime
}(s)+i\xi _{I}^{k\prime }(s)\}+\frac{i}{2}\epsilon _{ijk}\{\xi _{R}^{i}+i\xi
_{I}^{i}\}\{\xi _{R}^{j\prime }+i\xi _{I}^{j\prime }\}
\]%
or with, $\xi _{I}^{0}=v_{R}^{a}\xi _{I}^{b}\eta _{ab}$ as correct to first
order, and adding in the time components of $\xi ^{a}(s),$ we obtain a four
dimensional expression
\begin{equation}
Z^{a}(s)=\xi _{R}^{a}(s)+i\xi _{I}^{a}(s)-i(v_{R}^{c}\xi _{I}^{b}\eta
_{cb})[v_{R}^{a}(s)+iv_{I}^{a}(s)]+\frac{i}{2}\epsilon
_{.bcd}^{a}v_{R}^{d}\{\xi _{R}^{b}+i\xi _{I}^{b}\}\{v_{R}^{c}+iv_{I}^{c}\}
\label{Z^a}
\end{equation}%
that can be used to define relativistically the complex ($electric+i$ $%
magnetic$) dipole moments,
\begin{eqnarray}
D_{C}^{a}(s) &=&qZ^{a}(s)  \nonumber \\
&=&q[\xi _{R}^{a}(s)+i\xi _{I}^{a}(s)-iv_{R}^{c}\xi _{I}^{b}\eta
_{cb}(v_{R}^{a}(s)+iv_{I}^{a}(s))  \label{dipole from line} \\
&&+\frac{i}{2}\epsilon _{.bcd}^{a}v_{R}^{d}\{\xi _{R}^{b}+i\xi
_{I}^{b}\}\{v_{R}^{c}+iv_{I}^{c}\}].  \nonumber
\end{eqnarray}

Alternatively, assuming that we have been given the Maxwell field and hence
a known $Z^{k}(s)$, we can then solve Eq.(\ref{condition}) for the spatial
part of the world-line:
\begin{equation}
\xi ^{k}(s)=Z^{k}(s)+i\xi _{I}^{0}(s)Z^{k\prime }(s)-i\epsilon _{ijk}\frac{1%
}{2}Z^{i}(s)Z^{j\prime }(s).  \label{line.from.dipole}
\end{equation}

The time component of $\xi _{R}^{a}$ is determined from the unit
normalization while the time part of $\xi _{I}^{a}$ is determined, up to an
arbitrary constant, by an integration coming from the orthogonality
condition $v_{I}^{a}v_{R}^{b}\eta _{ab}=0.$ For example, in the rest frame,
i.e., when $v_{R}^{b}=(1,0,0,0),$ we would have that $v_{I}^{0}=0$ and that $%
\xi _{I}^{0}$ = constant.

{\bf Remark 11.} \textit{Though we do not have a simple argument for
it, it seems almost certain to us that this constant should be taken
to be zero.}
\\

{\bf Remark 12.} Note that the relations between the world-line,
$\xi _{R}^{a}(s),$ and the dipole moment $qZ(s),$ depend not only on
their instantaneous values but also on their derivatives. If we had
not used the severe approximation of terminating the $\lambda $
expansion and the slow motion assumption then the information about
their entire histories would appear in the relations.
\\

The physical meaning of $\lambda $, from Eq.(\ref{LAMBDA*})

\begin{equation}
\lambda (s,\zeta ,\overline{\zeta })=-\frac{\xi _{I}^{b\,}(s)l_{b}}{\xi
_{R}^{c\,\prime }(s)l_{c}}\approx -\sqrt{2}\xi _{I}^{b\,}(s)l_{b}
\label{LAMBDA***}
\end{equation}%
is that it carries the full information about the imaginary part of the
world-line or, modulo $q,$ about the \textit{magnetic dipole} \textit{moment}%
.

\section{Conclusion}

As we pointed out in the introduction, our interest here concerned two
related issues - first: the relationship between regular shear-free null
geodesics in Minkowski space, its associated CR structure (or CR function)
defined on $\mathfrak{I}^{+}$ and a complex world-line defined on complex
Minkowski space and second: the relationship between the Maxwell fields and
the decomposition of the CR function into real and imaginary parts. After a
canonical choice was made of $\tau ,$ the CR function $\tau =T(u,\zeta ,%
\overline{\zeta })$ and its inversion $u_{C}=X(\tau ,\zeta ,\overline{\zeta }%
)=\xi ^{b\,}(\tau )l_{b}(\zeta ,\overline{\zeta })$ were written, with $\tau
=s+i\lambda ,$ as

\begin{eqnarray*}
s &=&T_{R}(u,\zeta ,\overline{\zeta }) \\
\lambda &=&T_{I}(u,\zeta ,\overline{\zeta })
\end{eqnarray*}
and

\begin{eqnarray*}
u &=&\chi _{R}(s,\Lambda (s,\zeta ,\overline{\zeta }),\zeta ,\overline{\zeta
}). \\
\lambda &=&\Lambda (s,\zeta ,\overline{\zeta })
\end{eqnarray*}
with real $(u,s,\lambda ).$

We then saw that when the Maxwell field was used to determine the complex
world-line, the world-line, written as a function of the real $`s$', i.e.,
Eq.(\ref{wordline s}), was equivalent to knowledge of the electric and
magnetic dipole moment of the field.

The question then arises: can the ideas described here - the decomposition
of the CR function and its relationship to physical quantities - be extended
to general relativity.

It appears clear to us that this can be done. In particular, we should be
able to repeat the same basic arguments for all asymptotically flat Einstein
or Einstein-Maxwell space-times. Though it probably would be simpler to do
it for the algebraically special space-times containing a regular $\mathfrak{%
I}^{+},$ where we would again have regular shear-free null geodesics and
their associated CR structures and complex world-lines, we could also repeat
it for all asymptotically flat Einstein-Maxwell space-times\cite{footprints}%
. \ In the latter case, though there are, in general, no shear-free null
geodesic congruences, they can be replaced by the \textit{asymptotically}
shear-free null geodesic congruences which have most of the same relevant
properties as do the fully shear-free congruences. They also define a CR
structure, that can be realized on $\mathfrak{I}^{+},$ containing a complex
world-line (now defined on the associated H-space). Again the CR functions
can be decomposed into real and imaginary parts and the complex world-line
reinterpreted as a real world-line which carries with it a further real
structure - with both the real world-line and the real structure having
physical meaning.

Though this work is still to be completed, it seems very likely that the
real world-line can be interpreted as describing the motion of the center of
mass of the total gravitating system\cite{RTM} while the accompany real
structure contains or defines the spin-angular momentum of the system.

The calculations that are needed to confirm these conjectures are rather
long and detailed so that as a preliminary step, to get perspective, we are
first doing the calculations for the special case where the shear-free
congruence is also twist-free, a generalization of the Robinson-Trautman
case.

\section{Acknowledgments}

This material is based upon work (partially) supported by the National
Science Foundation under Grant No. PHY-0244513. Any opinions, findings, and
conclusions or recommendations expressed in this material are those of the
authors and do not necessarily reflect the views of the National Science.
E.T.N. thanks the NSF for this support. G.S.O. acknowledges the financial
support from Sistema Nacional de Investigadores (SNI-M\'{e}xico). C.K.
thanks CONICET and SECYTUNC for support.

\section{Appendix}

\subsection{The CR Structure}

An embedded CR manifold\cite{RP3,LNT,LNT2,LN,Taf,Tr} is a smooth real
submanifold embedded in \textbf{C}$^{n}.$ The embedding functions are
referred to as CR functions while the intrinsic description of the CR
manifold is given by, what is referred to as the CR structure, an
equivalence class (under a family of gauge transformations) of one-forms.
For us, we consider the real three-dimensional $\mathfrak{I}^{+}$ as
embedded in \textbf{C}$^{2}.$

The CR structure\cite{LNT,LNT2,LN,Taf,Tr} arises in the following manner. We
begin with Bondi coordinates $(u,\zeta ,\bar{\zeta})$ on $\mathfrak{I}^{+}$
and with a Bondi one-form basis $(n,l,m,\bar{m}).$ The one-form $n$ is the
dual to the tangent vector to the generators of $\mathfrak{I}^{+}$ and $l$
is dual to the vectors normal to the $u=constant$ slices of $\mathfrak{I}%
^{+},$ ($m,$ $\bar{m}$ are duals to the tangent vectors of the `slices', $%
u=constant$). We then perform a null rotation around $n$ of the form, Eq.(%
\ref{NULLROT^}),
\begin{equation}
l^{\ast }=l+\frac{L}{r}\bar{m}+\frac{\bar{L}}{r}m+O(r^{-2}),~~~~m^{\ast }=m+%
\frac{L}{r}n+O(r^{-2}),~~n^{\ast }=n  \label{nullRot}
\end{equation}%
where $L$ is a solution to the shear-free equation, Eq.(\ref{Shearfree}),
\[
\text{\dh }L+LL,_{u}=0.
\]%
We consider the (regular)\cite{CR} solutions $L=L(u,\zeta ,\bar{\zeta})$
depending on an arbitrary complex world-line, Eq.(\ref{L&u}). The resulting
dual one-forms on $\mathfrak{I}^{+}$ are (after a conformal rescaling of $m$%
)
\begin{eqnarray}
l^{\ast } &=&du-\frac{L}{1+\zeta \bar{\zeta}}d\zeta -\frac{\bar{L}}{1+\zeta
\bar{\zeta}}d\bar{\zeta},~~~  \label{one-forms} \\
~m^{\ast } &=&\frac{d\overline{{\zeta }}}{1+\zeta \bar{\zeta}},\qquad ~%
\overline{m}^{\ast }=\frac{d{\zeta }}{1+\zeta \bar{\zeta}}.  \nonumber
\end{eqnarray}%
The three one-forms are a representative set of one-forms (up to the gauge
freedom) that define the CR structure on$~\mathfrak{I}^{+}.$ The embedding
is given by two functions, $E^{i}(u,\zeta ,\bar{\zeta})=(E^{1},E^{2}),$ that
satisfy the CR equation

\begin{equation}
\text{\dh }E^{i}+L\partial _{u}E^{i}=0.  \label{CREq}
\end{equation}

Choosing ($\tau ,\widetilde{\zeta }$) as coordinates on \textbf{C}$^{2},$
two independent solutions of Eq.(\ref{CREq}), (see Eq.(\ref{Lofphi})) are
given by

\begin{eqnarray*}
\tau &=&T(u,\zeta ,\bar{\zeta})\equiv E^{1}(u,\zeta ,\bar{\zeta}) \\
\widetilde{\zeta } &=&\overline{\zeta }\equiv E^{2}.
\end{eqnarray*}%
We saw earlier that $\tau =T(u,\zeta ,\bar{\zeta})$ could be obtained as the
inverse function to $u=\xi ^{a}(\tau )l_{a}$ \ so that every complex
world-line defines a CR structure on $\mathfrak{I}^{+}$ while every
asymptotically vanishing Maxwell field picks out a unique CR structure.

\end{document}